Temporal and Semantic Effects on Multisensory Integration

Jean M. Vettel[a,b,c], Julia R. Green[d], Laurie Heller[e], Michael J. Tarr[e]

[a] U.S. Army Research Laboratory
[b] University of California, Santa Barbara
[c] University of Pennsylvania
[d] Brown University
[e] Carnegie Mellon University

Abstract

How do we integrate modality-specific perceptual information arising from the same physical event into a coherent percept? One possibility is that observers rely on information across perceptual modalities that shares temporal structure and/or semantic associations. To explore the contributions of these two factors in multisensory integration, we manipulated the temporal and semantic relationships between auditory and visual information produced by real-world events, such as paper tearing or cards being shuffled. We identified distinct neural substrates for integration based on temporal structure as compared to integration based on event semantics. Semantically incongruent events recruited left frontal regions, while temporally asynchronous events recruited right frontal cortices. At the same time, both forms of incongruence recruited subregions in the temporal, occipital, and lingual cortices. Finally, events that were both temporally and semantically congruent modulated activity in the parahippocampus and anterior temporal lobe. Taken together, these results indicate that low-level perceptual properties such as temporal synchrony and high-level knowledge such as semantics play a role in our coherent perceptual experience of physical events.

Introduction

Real-world events produce both auditory and visual signals that, together, comprise our perceptual experience of that event. At issue is how the brain integrates information from these independent perceptual channels to form coherent percepts. Low-level factors, such as common temporal and spatial structure (Meredith and Stein, 1993), and high-level factors, such as semantic labeling (Doehrmann and Naumer, 2008), both appear to influence integration. Here we examined the effects of both temporal synchrony and semantic congruency on the neural integration of audiovisual (AV) events.

Early research on multisensory integration revealed localized "heteromodal regions" (Beauchamp et al., 2004;Calvert et al., 2000), while more recent research has identified a distributed network of multisensory regions that includes primary sensory areas, posterior temporal cortex, posterior parietal areas, and inferior frontal areas (Driver and Noesselt, 2008;Kayser and Logothetis, 2007). Given such a wide array of potential integration sites, it is of interest to explore how different integration cues modulate and recruit this network.

Many studies have manipulated only a single factor. Studies of temporal congruency have manipulated the onset synchrony of simple stimuli (e.g., circles and tones) to identify brain regions sensitive to temporal coincidence between auditory and visual signals (Bischoff et al., 2007;Bushara et al., 2001). Studies of semantic congruency have paired static objects with their characteristic sounds (e.g., dog with barking, hammer with pounding) to identify brain regions sensitive to high-level associations (Belardinelli et al., 2004;Taylor et al., 2006). In contrast, investigating the interplay between temporal and semantic factors necessitates using stimuli in which information in both perceptual modalities unfolds over time.

Examples of dynamic stimuli can be found in studies of AV speech: a video of an articulating mouth paired with a spoken phoneme. Yet these studies often manipulate only a single factor – temporal synchrony (Jones and Callan, 2003;Miller and D'Esposito, 2005;Olson et al., 2002) or phonetic congruence (Ojanen et al., 2005;Skipper et al., 2007). A few studies have manipulated two cues simultaneously: temporal versus spatial lip-reading (Macaluso et al., 2004), semantic versus spatial object-sound matching (Sestieri et al., 2006), and semantic versus stimulus familiarity of animals and objects (Hein et al., 2007). Only one study manipulated semantic and temporal relationships, as in our present study, but the stimuli were printed letters and spoken phonemes, a pairing that may recruit neural substrates specialized for reading (van Atteveldt et al., 2007).

Our fMRI study examines the interplay of two cues to integration – temporal synchrony and semantic congruence – using real-world physical events. We filmed AV events, such as tapping a pencil, and edited them to create three multisensory conditions: *congruent* where both semantic and temporal information match across modalities (AVC), *temporally incongruent* where semantics is congruent but timing is asynchronous (AVTI), and *semantically incongruent* where both semantics *and* timing are incongruent (AVSI). Comparisons of the neural activity for these conditions implicate a network of functionally-distinct brain regions involved in multisensory integration. More broadly, our results indicate that both low-level temporal properties in the signal and high-level semantic knowledge help form our integrated experience of events.

Materials and Methods

*Participants.* Nineteen right-handed subjects (10 female/9 male; mean age: 23; range: 18-35) participated in the study, but data from four subjects (2 female/2 male) were discarded due to audio equipment failure during scanning. Subjects were paid for their participation and all provided written, informed consent consistent with and approved by Brown University IRB. All had normal or corrected-normal vision and hearing, and they reported that they were neurologically healthy with no contraindications for MRI scanning. The study was conducted at the Magnetic Resonance Facility at Brown University (Providence, RI).

*Stimuli.* A total of 40 AV movies were digitally filmed at standard definition for this study, including two exemplars of 20 unique real-world, physical events (e.g., typing on a keyboard, bouncing ball, jingling keys, pouring water, etc; for a complete list, please see Supplemental Document D1). In each movie, only the arm and hand of the actress were seen performing the event (Figure 1). The movies were edited using Final Cut Pro (Apple Co., Cupertino, CA) to be 2300 msec in duration and exported as 720x480 Quicktime™ (Apple Co., Cupertino, CA) movies with 16 bit, 44.1kHz stereo audio.

*Experimental Conditions.* Each of the 40 events was presented in five conditions, two unimodal, auditory only (A) and visual only (V), and three multisensory conditions – a *congruent* condition (AVC) and two incongruent conditions, *temporally incongruent* (AVTI) and *semantically incongruent* (AVSI). The congruent AV movies always showed audio and video from the same exemplar of a given event type. The AVSI movies showed the video from one event type (e.g., door knocking) simultaneously with the sound from a different event type (e.g., splashing water).

The AVTI movies showed audio and video from the same event type (e.g., door knocking) but the movie was taken from one of the two door knocking exemplars (door knocking B) and the sound was taken from the other exemplar of the same event type (door knocking A). Thus, both modalities should elicit the same semantic label ("door knocking") but the timing between the two modalities was asynchronous (Figure 1). The AVTI condition was always generated by pairing the video from one exemplar with the audio from the other exemplar of the same event type. For the AVSI condition, each event was randomly paired with a different physical event for each subject.

*Experimental Task.* Prior research has suggested that making an explicit congruency judgment can effect the obtained pattern of neural responses (van Atteveldt et al., 2007); instead, our subjects performed a simple stimulus location judgment, indicating whether the video or audio (different scans) was seen or heard on the right or left side. Each video was offset by 10% of the screen size (100 pixels) from the center fixation, and each sound was delayed by 500 microseconds in one ear so that it was lateralized towards the opposite ear (amplitude was held constant between the two ears so as not to introduce spurious differences in the neural signals between brain hemispheres). In all AV conditions, the audio and video were always offset in the same direction so as to be perceived as being in the same location (i.e., no spatial location incongruency). All subjects pressed the right button with their right index finger for right side and left button with left index finger for left side, thereby maintaining response congruency. Although the right/left location of the two modalities was always congruent, subjects only reported either the location of the video or the location of the audio so that the identical task could be used in the unimodal conditions. The modality used for the judgment alternated between scans and the order of the judgment was counterbalanced between subjects. As expected, accuracy was equally high for all five conditions across both modalities (96% +/- .01%).

*Experimental Procedure.* Before the scan, each subject practiced the location judgment task using movies from a different study (Zacks et al., 2006) that were not seen during the actual experiment. Subjects then studied a list of the 20 unique physical event types to appear in the experiment, performed a written recall task, and then read aloud the name of any event type not recalled. This procedure ensured that all events were familiar and identifiable during the scanning session. During the scanning session, high resolution anatomical MP-RAGE images were collected first, followed by four functional EPI scans using a fast event-related design. Each scan consisted of 20 unimodal trials (visual-only for scans judging video location and auditory-only for scans judging sound location), 10 AVC, 10 AVSI, and 10 AVTI multisensory trials. Thus, each scan had 50 experimental trials intermixed with 40 baseline fixation trials in stimulus orders that were optimized for GLT parameter estimation (AFNI's RSFgen, http://afni.nimh.nih.gov/afni/doc/howto/3). Each EPI scan lasted four and a half minutes.

*fMRI Details.* MRI data were acquired using a 3T Siemens TIM Trio scanner (Erlangen, Germany). Visual stimuli were presented through a rear projection system with a mirror mounted to the head coil, and audio stimuli presented through an Avotec SS-3100 Silent Scan audio system (Stuart, FL). Participants made behavioral responses with their left and right index fingers using the two outside buttons on a 4 button response pad (Mag Design & Engineering, Sunnyvale, CA). The high resolution MPRAGE scan included 160 interleaved slices with 1mm isotropic voxels in a 256x256 matrix with a TR=1900msec, TE=2.98msec, and flip angle=9degrees, and the T1-weighted echoplanar images (EPI) covered the whole brain with 48 interleaved slices, 3mm isotropic voxels, TR=3000msec, TE=28msec, and flip angle=90degrees. Each EPI scan consisted of 90 volumes. Prior to analysis in AFNI (Cox, 1996), the functional data were pre-processed in AFNI to correct for slice timing differences and 3-D head movement, smoothed with a FWHM kernel of 1 voxel (3mm), and normalized to allow the beta coefficients to reflect percent signal change by dividing each time series by its mean.

*fMRI Data Analysis*

Prior to warping the data to the standard atlas (Talairach and Tournoux, 1988), the data for each individual was fit with a general linear model (GLM) to estimate the activity at each voxel using a regressor for each of the five experimental conditions (two unimodal and three multisensory conditions) as well as regressors of no interest to account for second-order polynomial trends. A gamma hemodynamic response function was assumed. The output of each individual participant's GLM was then warped to the standard atlas and analyzed with a mixed-effects group ANOVA that included the three pairwise statistical contrasts of the parameter estimates for the multisensory conditions: AVC minus AVTI (temporal processing), AVTI minus AVSI (semantic processing), and AVC minus AVSI (congruency processing). In addition, a conjunction analysis identified integration regions with voxels more active in the AVC condition than either unimodal condition, with the additional requirement that unimodal condition must also be significantly greater than baseline, that is, $0 < A_{only} < AVC > V_{only} > 0$ (Beauchamp, 2005).

All statistical maps were thresholded at $p < 0.05$ with a requirement that all regions of interest (ROIs) have at least 24 voxels (connected by 3mm, i.e., one side of a voxel) to ensure an alpha level of 0.05 (AFNI's AlphaSim program written by B. Douglas Ward, run with 5000 Monte Carlo simulations).

Results

*Unimodal Regions*

As expected, a statistical contrast of auditory-only ($A_{only}$) versus visual-only ($V_{only}$) revealed extensive and separable neural activation for both unimodal conditions. Large clusters of activation for $A_{only}$ were found in primary auditory regions as well as the surrounding superior temporal gyrus, insula, supramarginal gyrus, and medial frontal gyrus. Widespread activation for $V_{only}$ was observed throughout the posterior areas in the brain, including primary visual areas, fusiform gyrus, lingual gyrus, cuneus/precuneus, and parietal cortex.

*AV Contrast: Integration Regions-of-Interest (ROIs)*

As discussed in the Methods section, we adhered to a functional criteria for integration defined by prior research (Beauchamp, 2005); that is, integration regions were identified by a conjunction analysis of brain regions that were more active in the AVC condition than either unimodal condition with the additional constraint that both unimodal conditions were greater than the baseline activity.

Four regions met this criteria: 1) a medial region in the posterior cingulate gyrus (pCing.G); 2) a posterior region in left middle temporal gyrus (L MTG); 3) a region slightly anterior to L MTG in the middle temporal gyrus that also extended upwards to the superior temporal sulcus (L STG/MTG); and 4) a parietal region in the left hemisphere.

These ROIs are shown in Figure 2A with further details presented in Table 1. Note that the two incongruent conditions were *not* used to identify the integration ROIs; a *t*-test between these two conditions showed no significant difference, indicating that the ROIs associated with functionally-defined integration did not differentiate between the two incongruent conditions (pCing.G $t(14)=0.63$, $p=0.54$; L MTG $t(14)=0.74$, $p=0.47$; L STG/MTG $t(14)=0.42$, $p=0.68$; L Parietal $t(14)=0.15$, $p=0.88$).

*AV Contrast: Temporal Processing*

The contrast between the AVC and AVTI conditions identifies brain regions that are sensitive to whether the timing relationship between the visual and auditory signals is synchronous for events that are semantically congruent; in other words, both the visual and auditory modalities are labeled as door knocking but the timing of the knock in the visual domain is not in synchrony with the sound of the door knock in the auditory domain.

Six regions involved in temporal processing are depicted in Figure 3 with further details presented in Table 1: an anterior (maroon) and a posterior (orange) region in the right middle temporal gyrus, bilateral regions (blue and red) in the lingual gyrus, a region that spans from the superior edge of the inferior frontal gyrus up through the medial frontal gyrus (pink) in the right hemisphere, and a medial region in the superior frontal gyrus (purple).

Two additional regions were identified in posterior cingulate gyrus and parietal cortex. These two are depicted in Figure 2B (green) to illustrate their spatial relationship to the integration regions identified in these same anatomical regions but with distinct activation clusters.

*AV Contrast: Semantic Processing*

The contrast of AVTI and AVSI identifies regions that are sensitive to the semantic labels that arise as a result of event identification between the visual and auditory domains. Critically, for these conditions the temporal structure of the visual and auditory signals is incongruent in both cases. In other words, the visual and auditory modalities are not temporally in synchrony for either condition, but in the AVTI condition, both modalities should be labeled the same (e.g., "door knocking"), while in the AVSI condition, the auditory signal should prompt one label and the visual signal should prompt a different label (e.g., "door knocking" and "water splashing").

Eight regions involved in semantic processing are depicted in Figure 4 with further details presented in Table 1: a region in left middle occipital gyrus (dark blue), a region in right lingual gyrus (dark green; lateral and superior to temporal processing Ling.G), a more lateral and anterior region in right middle occipital gyrus (light blue), a region in right middle temporal gyrus (orange; superior and posterior to the temporal processing R MTG region), a region that spans from the superior edge of the inferior frontal gyrus up through the medial frontal gyrus in the left hemisphere (pink), a region slightly superior in the medial frontal gyrus that extends in to superior frontal gyrus (purple; lateral to the temporal processing MFG/SFG region), and a region in the caudate (maroon).

Three separate regions in the cerebellum (not pictured), one region in middle cingulate gyrus (not pictured), one region in posterior cingulate gyrus, and two regions in parietal cortex were identified. The posterior cingulate and one of the parietal regions (row 14 in Table 1) are depicted in Figure 2B (orange) to illustrate their spatial relationship to the integration regions as well as the other AV contrasts regions. The not-pictured parietal ROI is just superior to the pictured one, and it also spans both hemispheres.

*AV Contrast: Congruency Processing*

The contrast of AVC and AVSI identifies brain regions that are sensitive to overall congruence between the visual and auditory modalities; in other words, a movie of water splashing is asynchronous with the sound of door knocking in that both the temporal structure and the semantic labeling provide information that the two domains are incongruent.

Seven regions involved with congruency processing are depicted in Figure 5 with further details presented in Table 1: a region in left middle temporal gyrus (orange; posterior and superior to the integration regions in MTG), a region in left medial frontal gyrus (light blue; posterior to the semantic processing L MFG region), a region in medial superior frontal gyrus (yellow; overlaps with temporal processing medial SFG), bilateral regions that span from the superior edge of the inferior frontal gyrus up through the medial frontal gyrus (purple and pink; the purple overlaps with the semantic processing L IFG/MFG region, and the pink overlaps with the temporally processing R IFG/MFG region), a region in the parahippocampus (dark blue), and a region in the caudate (maroon; overlaps with the semantic processing caudate region).

One region in the cerebellum (not pictured), one region in posterior cingulate gyrus, and one in parietal cortex were identified. The latter two are depicted in Figure 2B (teal blue) to illustrate their spatial relationship to the integration regions as well as the other AV Contrasts regions.

*Posterior Cingulate*

In each of the four AV contrasts, regions were found in the posterior cingulate, and these are depicted in Figure 2B and listed in Table 1. The most extensive activity arose from the Congruency Processing comparison (light blue). The inferior portion of this region overlapped with the Semantic Processing ROI (orange). This overlap region (pink) may reflect computations concerning semantic relationships since both AV contrasts compare a semantically congruent condition (AVC and AVTI) to the AVSI condition. This overlap region seems insensitive to the temporal relationship, as one contrast is temporally synchronous (AVC) and one asynchronous (AVTI). Although this region is adjacent to the Integration ROI, no voxels overlap.

The superior portion of the Congruency Processing ROI (light blue) overlaps with the Temporal Processing ROI (green), and this overlap region (maroon) may reflect sensitivity to the temporal relationships in multisensory stimuli since both of these AV contrasts compare the AVC condition, which is temporally synchronous between modalities, with two temporally incongruent conditions, that is, where the modalities are temporally asynchronous (AVTI and AVSI). In contrast, the overlap region appears insensitive to semantic relationships since one contrast is semantically congruent (AVTI) and the other is incongruent (AVSI). This activity is superior to the Integration ROI.

*Parietal Cortex*

As in the posterior cingulate, parietal regions were found in each of the three AV contrasts: these are depicted in Figure 2B and listed in Table 1. The integration ROI is inferior and lateral to the other parietal regions. The integration ROI is located within the left inferior parietal lobule, while the other ROIs are superior and located in the precuneus, very close to the midline, and extending upwards in the superior parietal lobule.

The Congruency Processing ROI (light blue) and Temporal Processing ROI (green) have a high percentage of overlap (maroon), and as described in the previous section on the Posterior Cingulate, the overlap of these two AV Contrast maps may identify a region sensitive to the timing relationship between modalities.

Superior to all of the other AV Contrast ROIs, the Semantic Processing ROI (orange) identifies two regions in the parietal cortex, one directly superior to the other. The more inferior of the two is shown in Figure 2B. Unlike in the posterior cingulate, there is no overlap of this ROI with the Congruency Processing ROI (light blue) to identify a region that may be more sensitive to the semantic relationship between modalities (that is, no pink voxels).

*Frontal Cortex*

A frontal ROI was identified in all three AV contrasts, and as shown in Figure 2B, there is once again overlap among the regions. In the right hemisphere, the Congruency Processing ROI (light blue) overlaps with the Temporal Processing ROI (green), and this overlap region (maroon) may reflect sensitivity to the temporal relationships. Conversely in the left hemisphere, the Congruency Processing ROI (light blue) overlaps with the Semantic Processing ROI (orange), suggesting a region (pink voxels) sensitive to the semantic relationship between modalities. Thus, the congruent AVC condition identifies bilateral regions, but there appears to be some hemispheric specialization for different cue types that is revealed only when the two integration cues are pitted against one another in the AVIT and the AVIS conditions.

*Resting State Network and the AV Contrasts*

This series of AV contrasts also revealed regions within the anterior portion of the medial frontal gyrus and anterior cingulate, regions thought to be part of the resting state network and characterized by deactivations that are largely task independent (Raichle et al., 2001). Consistent with this classification, extracting the ROI means from these anterior regions revealed that each condition in the contrast was deactivating the region. Other studies have found similar patterns of deactivation within this resting state network (Beauchamp, 2005;Laurienti et al., 2002) and excluded them from further analysis. Similarly, in our analysis these regions are excluded from further discussion though details are listed in Table 1.

Discussion

Our study investigated how semantic and temporal congruency interact and modulate the brain regions consistently identified by prior research on multisensory integration, including primary sensory areas, posterior temporal, posterior parietal, and inferior frontal regions (Driver and Noesselt, 2008;Kayser and Logothetis, 2007). Using pairwise comparisons between our AV conditions, we successfully identified ROIs in all of these regions, and our study suggests how this network of brain regions may be differentially recruited to support specific processing of the semantic and/or temporal relationship between modalities. In particular, one novel and unexpected result is the hemispheric difference in the frontal cortex where the right may be more sensitive to timing relationships and the left to semantic relationships. Interestingly, none of the four functionally-defined integration regions were modulated by congruency.

Because we explicitly examine the interplay between temporal synchrony and semantic congruency across modalities, we are better able to isolate specific neural subregions that may support a particular type of cue integration, rather than generic integration effects. That is, by identifying the separable components of the multisensory processing network, our study advances the understanding of how modality-specific information is bound into coherent event percepts. Consequently, our discussion will focus on the degree of overlap between AV conditions. However, before we address what these overlap analyses reveal about multisensory integration, we first review the functionally-defined approach to localizing integration regions.

*Functionally-defined Integration ROIs*

Four integration regions were identified that showed higher activation for the AVC condition than either of the unimodal conditions, with the constraint that both unimodal conditions were greater than baseline ($0 < A_{only} < AVC > V_{only} > 0$; (Beauchamp, 2005)): two left posterior temporal regions, one region near the left intraparietal sulcus, and one medial region in the posterior cingulate.

Based on earlier multisensory research (Beauchamp et al., 2004;Calvert, 2001), many recent studies have focused on the posterior STS/MTG as a site of integration, and often as the source of feedback to other regions in the multisensory network (van Atteveldt et al., 2007); however, several recent reviews have reported the absence of congruency effects within pSTS/MTG (Hocking and Price, 2008;Doehrmann et al., 2008;Hein et al., 2007) and other studies fail to find any multisensory effects in this brain area (Belardinelli et al., 2004;Bushara et al., 2001) or find congruency effects in anterior temporal regions instead (van Atteveldt et al., 2004). Our two temporal integration regions are within 3mm of the two average locations identified in Hocking & Price (2008; +/-50, -52, 8 & +/-50, -56, 4), and neither temporal nor semantic congruency modulated the region (Figure 2A). Furthermore, Hocking & Price (2008) report that there is no significant difference in activation between intramodal (two auditory or two visual stimuli) and cross-modal (auditory and visual) trials. Thus, the pSTS/MTG region may not process whether information across multiple modalities should be integrated into a common percept; instead, the pSTS/MTG may access amodal representations, as suggested by Hocking & Price, or serve as a gateway between the frontal and medial temporal lobes whose function varies based on task-dependent connections (Hein and Knight, 2008).

The parietal cortex is consistently activated in multisensory studies (Amedi et al., 2005;Driver and Noesselt, 2008), and our integration ROI matches the commonly activated region near the intraparietal sulcus. However, the locus of our semantic and temporal congruency effects lie superior and posterior in the precuneus. Similar regions have been found in numerous studies using stimuli as diverse as vertical bars and beeps (Bushara et al., 2003), animals and novel objects (Hein and Knight, 2008), and articulating mouths (Miller and D'Esposito, 2005;Ojanen et al., 2005). This area appears to be recruited when the two modalities are incongruent (semantically incongruent in Hein et al; phonetically incongruent in Ojannen et al; temporally incongruent in Miller & D'Esposito). Bushara et al (2003) found similar regions that were more active when the audio and visual stimuli were not bound into a coherent percept, which is likely to have occurred in our conditions that activate this region. Finally, parietal regions are indentified when studies have manipulated the spatial congruency of AV stimuli, finding that this region responds to spatially congruent stimuli (Saito et al., 2005;Sestieri et al., 2006), which is true for every AV condition in our study (and true for the studies cited above). Thus, the precuneus may be differentiating when modalities are spatially congruent but yet still in conflict (be it a semantic, phonetic, or temporal conflict).

To our knowledge, no other study has found an integration region in the posterior cingulate, but this likely reflects our novel stimulus set of environmental events (not the typical tools, animals, or speech). Consistent with this interpretation, Lewis and colleagues (2004) investigated brain regions recruited for recognizing environmental sounds (compared to unrecognizable reversed versions of each sound), and identified several regions including caudate, posterior cingulate, posterior MTG, and IFG that mirror our results. Similar to the parietal cortex, the posterior cingulate integration region did not show congruency effects but adjacent (but not overlapping) regions were modulated by temporal and/or semantic congruency. Future research is needed to delineate the role of these differentiated effects, both within the posterior cingulate and the precuneus.

*Overlap of Regions Among the 3 AV Comparisons*

Looking beyond the functionally-defined integration regions, the pairwise comparisons of the three multisensory conditions in the whole-brain analysis identified regions that may process a particular integration cue, semantics or timing. In particular, regions identified in two of the three AV comparisons that overlap with one another may indicate a sensitivity to one of the two integration cues that is independent of the other.

The overlap of regions in the Semantic Processing and Congruency Processing comparisons suggest a role in differentiating the semantic relationship between modalities, as both of these contrasts compare conditions with congruent semantics across modalities to the condition in which semantics are incongruent (mixing whether the timing relationship is synchronous or not). Consequently, these overlap regions suggest a sensitivity to semantics that is independent of the temporal relationship between modalities. These contrasts reveal overlap in the left IFG/MFG region. Left lateralized effects are consistent with the well-known language dominance in the left hemisphere, as well as previous multisensory research on lip reading

(Paulesu et al., 2003) and events/tools (Lewis et al., 2005;Doehrmann et al., 2008). Similarly, research using amodal semantic priming identified a similar ROI (Buckner et al., 2000). The only other region to overlap in these two contrasts is in the inferior portion of the posterior cingulate, and as discussed in the previous section, this region may be specific to the stimulus class of environmental events (Lewis et al., 2004).

Conversely, the overlap of regions in the Temporal Processing and Congruency Processing comparisons suggests a role in differentiating the temporal relationship between modalities, as both of these contrasts compare the condition that is temporally synchronous across modalities with the conditions in which the timing relationship is asynchronous (ignoring whether the semantic relationship is congruent or not). Thus, the overlap of these regions suggests a preference for timing relations that is independent of semantics. Interestingly, one of the overlap regions is in the right IFG/MFG, a region that is bilateral to the overlap region in left IFG/MFG identified for semantic processing. Thus, our study suggests a hemispheric difference in the frontal cortex, with semantics focused in the left hemisphere and temporal relationships in the right hemisphere. This novel result arises from the addition of the AVTI condition in our study, as prior research has typically looked at semantic congruency with conditions equivalent to our Congruency Processing (AVC-AVSI) comparison, finding bilateral activation in frontal cortices (Beauchamp et al., 2004;Naumer et al., 2008;Taylor et al., 2006;van Atteveldt et al., 2007). However, two studies that looked at onset synchrony of simple stimuli (e.g., circles/tones) have reported right hemisphere dominance for temporal synchrony (Bushara et al., 2001;Calvert et al., 2001). Finally, in addition to the left MFG/IFG region, two additional overlap regions sensitive to temporal relationships were found in the superior portion of the posterior cingulate and the medial region of the precuneus, both of which are discussed in the previous section.

*Summary*

Our study was designed to disentangle some of the factors that contribute to the integration of perceptual signals arising from multisensory events. More specifically, we were interested in exploring the neural substrates recruited by integration cues arising from both low-level information in the signal (e.g., common temporal structure) and high-level knowledge (e.g., the semantic labels applied to perceptual signals). To address this question we developed a unique stimulus set of movies showing real-world physical events that unfold over time. These movies allowed us to manipulate congruency between the temporal structure of the auditory and visual signals independently from the semantic labeling of the same auditory and visual signals. As a result, our study effectively isolated the separable effects of both signal-based and high-level cues and, in doing so, identifies the neural substrates independently and commonly recruited by these two contributors to the process of multisensory integration.

Acknowledgements: The authors would like to thank Emily Myers for analysis advice and insightful discussions as well as Lynn Fanella and Mike Worden for help with data collection and analysis. This research was supported by NSF Award #0339122, by the Temporal Dynamics of Learning Center (NSF Science of Learning Center SBE-0542013), by the Perceptual Expertise Network (#15573-S6), a collaborative award from James S. McDonnell Foundation, and through the generosity of The Ittleson Foundation. The MRI system used in the study was purchased in part by an MRI grant from the NSF. JMV was supported by a Department of Defense SMART Graduate Fellowship. The funders had no role in study design, data collection and analysis, decision to publish, or preparation of the manuscript.

Tables

**Table 1 ROI Location, Size, and Peak Coordinates for all AV Contrasts**

| Row Number & ROI Anatomical Location | Hemi. | # of Voxels | Integration ROIs | Con – Temp.I | Temp.I – Sem.I | Con – Sem.I | Overlap | Peak (Tailarch) x | y | z |
|---|---|---|---|---|---|---|---|---|---|---|
| 1 Middle/Superior Temporal | L | 29 | * | | | | - | -43.5 | -43.5 | 14.5 |
| 2 Middle Temporal (posterior) | L | 20 | * | | | | - | -52.5 | -64.5 | -0.5 |
| 3 Middle Temporal (posterior) | R | 27 | | [] | | | - | 61.5 | -46.5 | -6.5 |
| 4 Middle Temporal (posterior) | R | 28 | | | <> | | - | 55.5 | -61.5 | 17.5 |
| 5 Middle Temporal (posterior) | L | 24 | | | | /\ | - | -49.5 | -73.5 | 23.5 |
| 6 Middle Temporal (anterior) | R | 27 | | [] | | | - | 58.5 | 7.5 | -18.5 |
| 7 Cingulate (posterior) | R | 35 | * | | | | - | 1.5 | -58.5 | 11.5 |
| 8 Cingulate (posterior) | L/R | 117 | | [] | | | 10 | -10.5 | -49.5 | 26.5 |
| 9 Cingulate (posterior) | L | 31 | | | <> | | 10 | -1.5 | -46.5 | 8.5 |
| 10 Cingulate (posterior) | L/R | 418 | | | | /\ | 8 & 9 | -1.5 | -31.5 | 8.5 |
| 11 Cingulate (middle) | L | 24 | | | <> | | - | -16.5 | -25.5 | 29.5 |
| 12 Parietal | L | 12 | * | | | | - | -34.5 | -46.5 | 44.5 |
| 13 Parietal | L | 32 | | [] | | | 16 | -10.5 | -73.5 | 56.5 |
| 14 Parietal | L/R | 33 | | | <> | | - | -1.5 | -61.5 | 62.5 |
| 15 Parietal | L/R | 25 | | | <> | | - | -4.5 | -52.5 | 65.5 |
| 16 Parietal | L | 25 | | | | /\ | 13 | -4.5 | -61.5 | 47.5 |
| 17 Lingual | L | 51 | | [] | | | - | 4.5 | -91.5 | -6.5 |
| 18 Lingual | R | 35 | | [] | | | - | 16.5 | -103.5 | -9.5 |
| 19 Lingual | R | 62 | | | <> | | - | 13.5 | -100.5 | -0.5 |
| 20 Middle Occipital | L | 28 | | | <> | | - | -25.5 | -88.5 | 2.5 |
| 21 Middle Occipital | R | 25 | | | <> | | - | 46.5 | -67.5 | -3.5 |
| 22 Parahippocampus | R | 30 | | | | /\ | - | 22.5 | -19.5 | -6.5 |
| 23 Inferior/Middle Frontal | R | 68 | | [] | | | 24 | 46.5 | 10.5 | 47.5 |

| # | ROI | Hemi. | # of Voxels | Con | Temp.I | Sem.I | Overlap | x | y | z |
|---|---|---|---|---|---|---|---|---|---|---|
| 24 | Inferior/Middle Frontal | R | 56 | | | /\ | 23 | 43.5 | 7.5 | 41.5 |
| 25 | Inferior/Middle Frontal | L | 32 | | <> | | 26 | -43.5 | 10.5 | 35.5 |
| 26 | Inferior/Middle Frontal | L | 70 | | | /\ | 25 | -52.5 | 10.5 | 35.5 |
| 27 | Middle Frontal | L | 25 | | | /\ | - | -46.5 | 4.5 | 47.5 |
| 28 | Middle/Superior Frontal | L | 28 | | <> | | - | -28.5 | 19.5 | 50.5 |
| 29 | Superior Frontal | L/R | 51 | [] | | | 30 | -1.5 | 19.5 | 50.5 |
| 30 | Superior Frontal | L/R | 31 | | | /\ | 29 | 1.5 | 13.5 | 50.5 |
| 31 | Caudate | L | 26 | | <> | | 32 | -4.5 | 10.5 | -0.5 |
| 32 | Caudate | L | 40 | | | /\ | 31 | 1.5 | 7.5 | -0.5 |
| 33 | Cerebellum | L/R | 34 | | <> | | - | 22.5 | -31.5 | -33.5 |
| 34 | Cerebellum | R | 27 | | <> | | - | 22.5 | -55.5 | -21.5 |
| 35 | Cerebellum | L/R | 77 | | <> | | 36 | 1.5 | -64.5 | -33.5 |
| 36 | Cerebellum | L/R | 34 | | | /\ | 35 | 4.5 | -61.5 | -18.5 |
| 37 | Medial Frontal | L/R | 60 | [] | | | 40 | 4.5 | 58.5 | 11.5 |
| 38 | Anterior Cingulate | R | 29 | [] | | | 40 | -4.5 | 40.5 | -12.5 |
| 39 | Anterior Cingulate | L | 29 | | <> | | 40 | -1.5 | 43.5 | 5.5 |
| 40 | Medial Frontal/Ant. Cing. | L/R | 599 | | | /\ | 37-39 | 4.5 | 67.5 | 17.5 |

a. All regions were identified with a $p < .05$, minimum cluster size of 24 voxels, alpha .05
b. Abbreviations: ROI = Region-Of-Interest; Hemi. = Hemisphere; # of Voxels = Number of 3mm voxels in region; Con = Congruent; Temp.I = Temporally Incongruent; Sem.I = Semantically Incongruent
c. Overlap column lists the row number of the ROI that overlaps

Figure Legends

**Figure 1** *Experimental Design*
On the left, two exemplars are shown for two events, door knocking and water splashing. The audio track for each exemplar reveals the different temporal characteristics between the two exemplars. The right table provides an example of how the five different experimental conditions are constructed: the Audio only condition presents just the audio track, the Visual only condition presents just the video track, the AVC (congruent) condition presents the audio and video from same exemplar, the AVTI (temporally incongruent) condition presents the audio from exemplar A and the video from exemplar B, and the AVSI (semantically incongruent) condition presents the video from water splashing and the audio from door knocking.

**Figure 2** *Functionally-defined Integration ROI and Overlap of AV Contrast Regions*
(A) The four functionally-defined Integration ROIs are pictured on the brain from one subject in the study with the coordinates of the crosshairs listed below the image. The ROIs were defined on the group data, but under each region, the mean percent signal change for each ROI in each individual subject is plotted for all five experimental conditions ($A_{only}$ = Auditory only; $V_{only}$ = Visual only; AVC = congruent; AVSI = semantically incongruent; AVTI = temporally incongruent). The error bars reflect standard error of the means between subjects. A *t*-test was performed on the means for the two incongruent conditions, and the n.s. denotes each test was not significant (see Results, Integration ROIs).
(B) Group ROIs in the parietal, posterior cingulate, and frontal regions identified in the three pairwise AV contrasts are pictured: regions identified in the Semantic Processing comparison (AVTI-AVSI) are shown in orange; regions from Congruency Processing (AVC-AVSI) are shown in light blue; and regions from Temporal Processing (AVC-AVTI) in green. This figure highlights the overlap of the regions identified in these contrasts, with pink voxels identifying overlap between the Semantic Processing and Congruency Processing contrasts (suggesting the region is sensitive to semantics) and maroon voxels showing overlap between the Congruency Processing and Temporal Processing Contrasts (suggesting the region is sensitive to timing; see Results for more details). In addition, the functionally-defined integration regions in parietal and posterior cingulate cortices are shown in purple.

**Figure 3** *AV Contrast: Temporal Processing*
The regions identified by the AV contrast of AVC (congruent) and AVTI (temporally incongruent) are plotted in six different colors. The bar graphs on the right show the mean percent signal change in each group-defined ROI from each individual subject (bars are between-subject error): green for AVC and yellow for AVTI. The color of the bar graph label matches the color of the corresponding voxels. Abbreviations: R = right; L = left; a = anterior; p = posterior; MTG = middle temporal gyrus; Ling.G = lingual gyrus; IFG = inferior frontal gyrus; MFG = middle frontal gyrus; SFG = superior frontal gyrus

**Figure 4** *AV Contrast: Semantic Processing*
The regions identified by the AV contrast of AVTI (temporally incongruent) and AVSI (semantically incongruent) are plotted in eight different colors. The bar graphs on the right show the mean percent signal change in each group-defined ROI from each individual subject (bars are between-subject error): yellow for AVTI and red for AVSI. The color of the bar graph label matches the color of the corresponding voxels. Abbreviations: R = right; L = left; p = posterior; Mid. Occ. = middle occipital gyrus; Ling.G = lingual gyrus; MTG = middle temporal gyrus; IFG = inferior frontal gyrus; MFG = middle frontal gyrus; SFG = superior frontal gyrus; Cing.G = cingulate gyrus

**Figure 5** *AV Contrast: Congruency Processing*
The regions identified by the AV contrast of AVC (congruent) and AVSI (semantically incongruent) are plotted in seven different colors. The bar graphs on the right show the mean percent signal change in each group-defined ROI from each individual subject (bars are between-subject error): green for AVC and red for AVSI. The color of the bar graph label matches the color of the corresponding voxels. Abbreviations: R = right; L = left; MTG = middle temporal gyrus; MFG = middle frontal gyrus; SFG = superior frontal gyrus; IFG = inferior frontal gyrus; Parahipp. = parahippocampus

|              | Door Knock | Water Splash |
|--------------|------------|--------------|
| Exemplar A   | 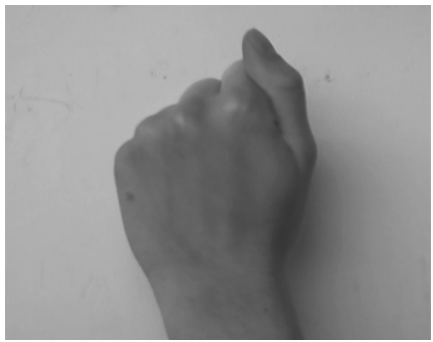<br>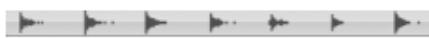 | 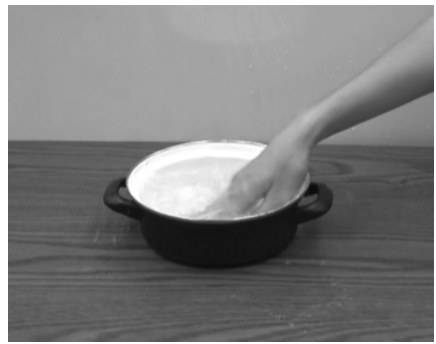<br>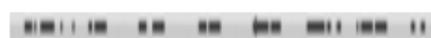 |
| Exemplar B   | 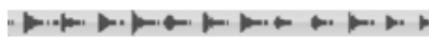<br>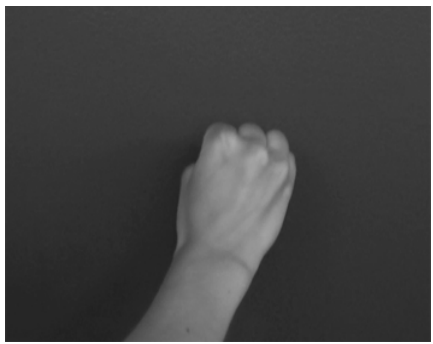 | 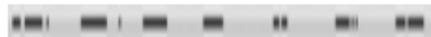<br>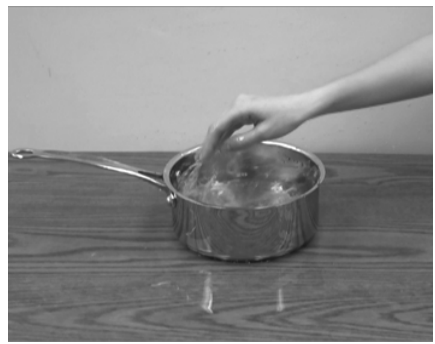 |

|             | KnockA | KnockA | KnockB | SplashA |
|-------------|--------|--------|--------|---------|
| Audio only  | ○      |        |        |         |
| Visual only |        | ○      |        |         |
| AVC         | ○      | ○      |        |         |
| AVTI        | ○      |        | ○      |         |
| AVSI        | ○      |        |        | ○       |

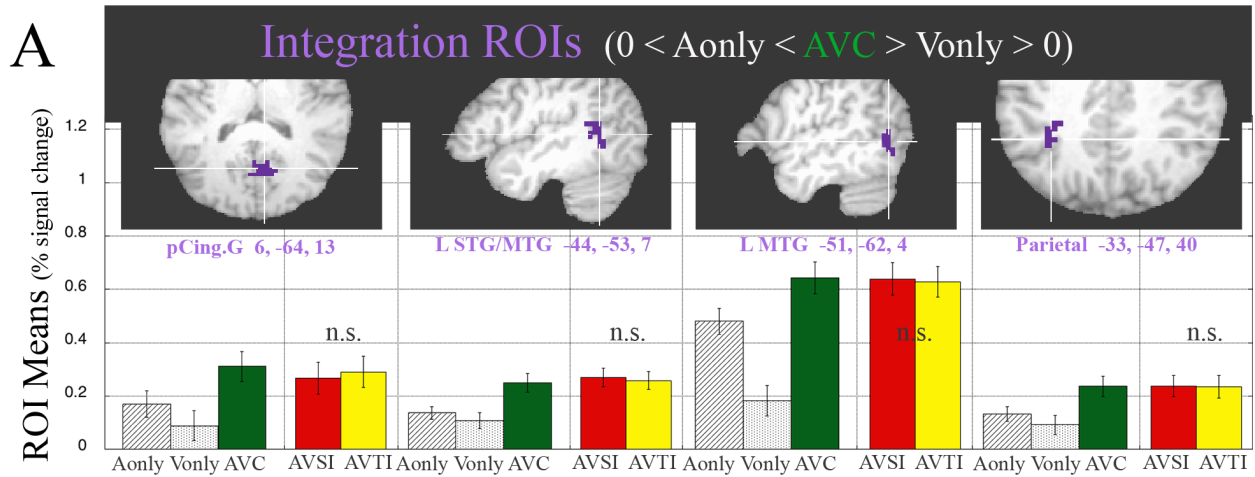

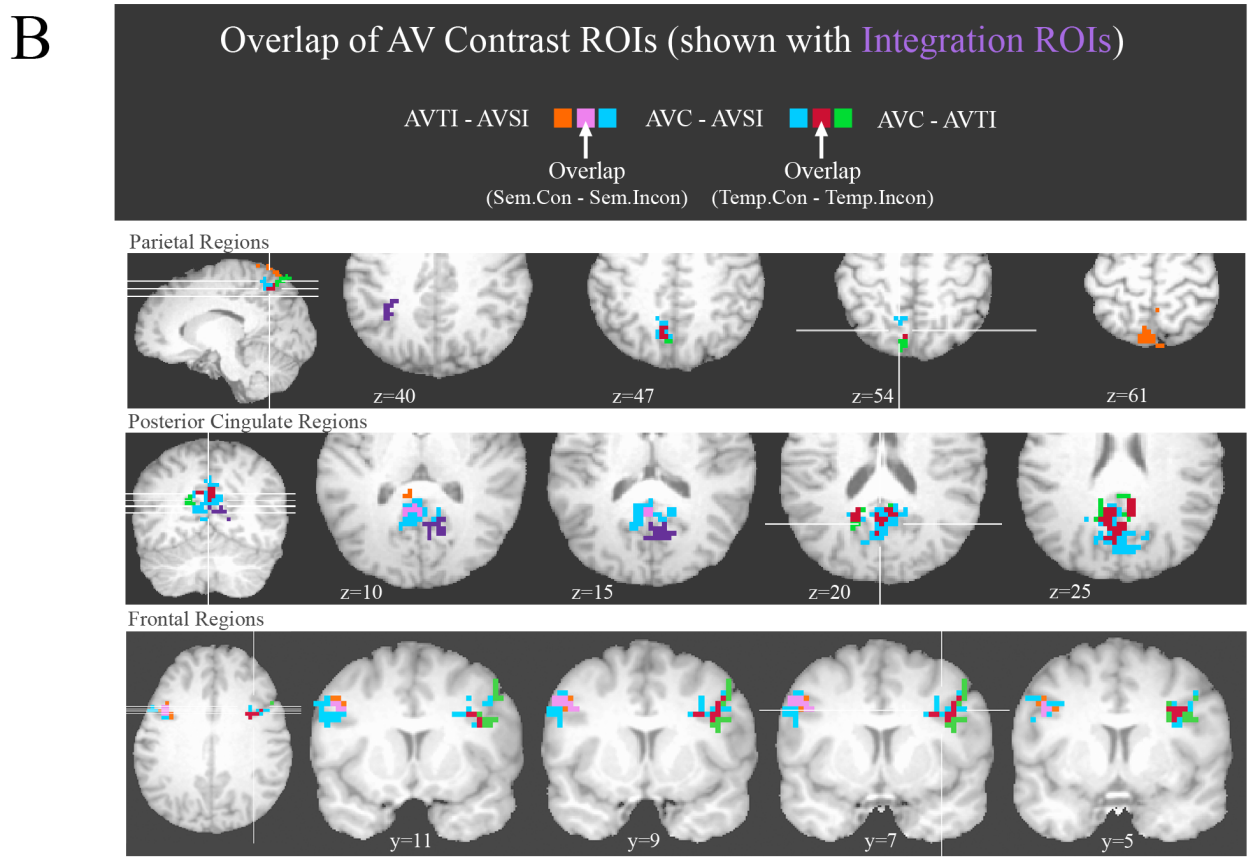

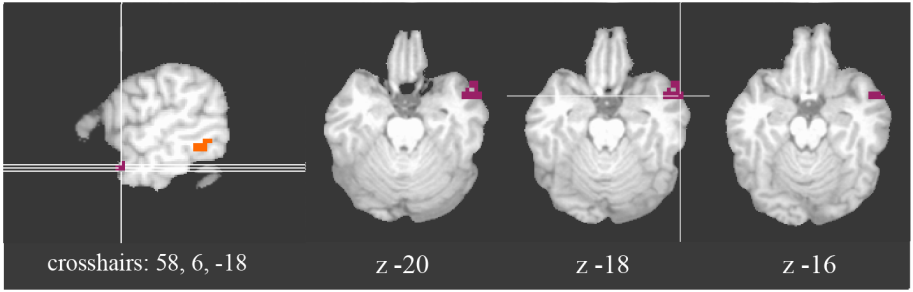
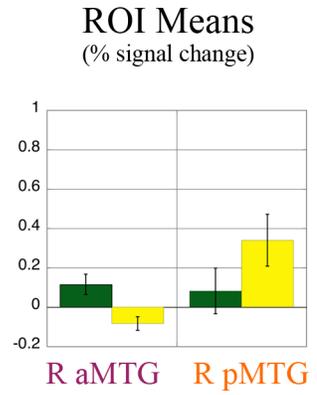
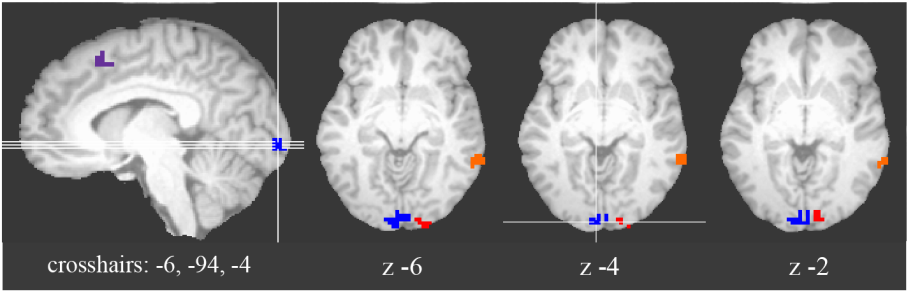
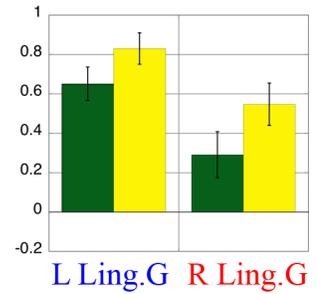
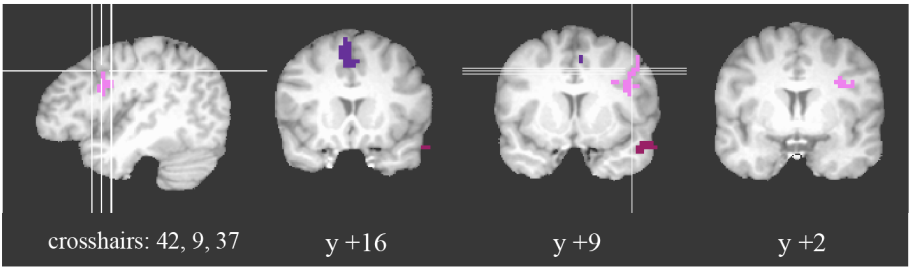
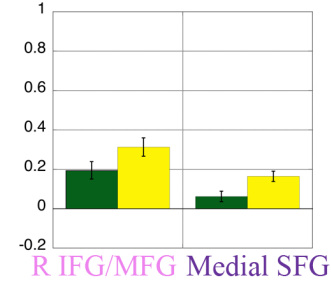

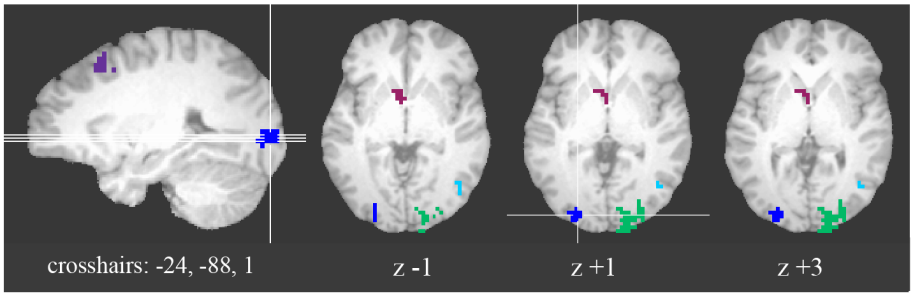
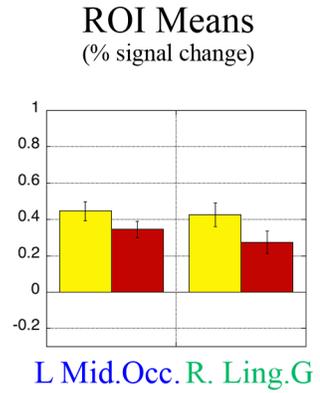
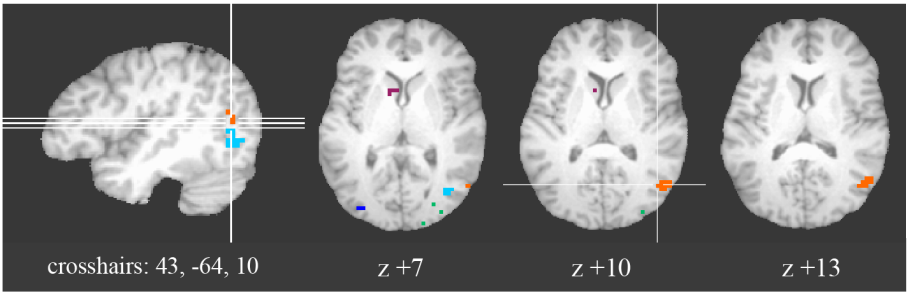
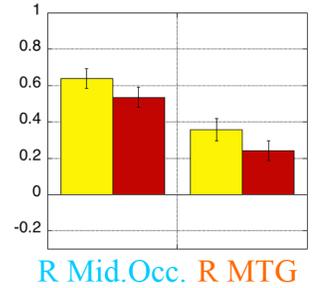
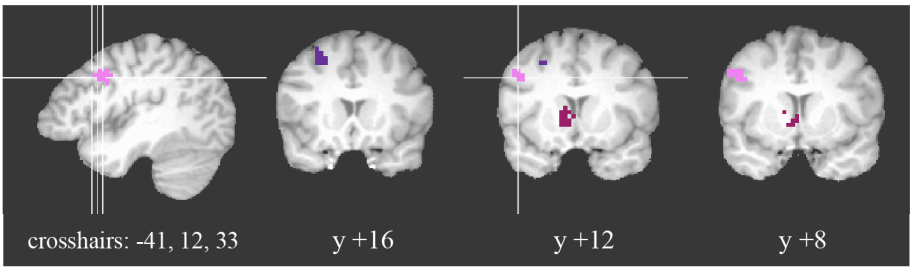
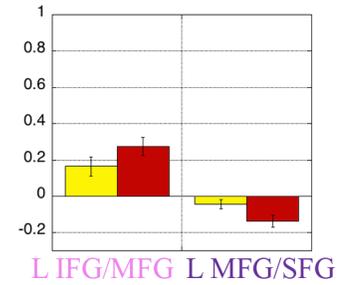
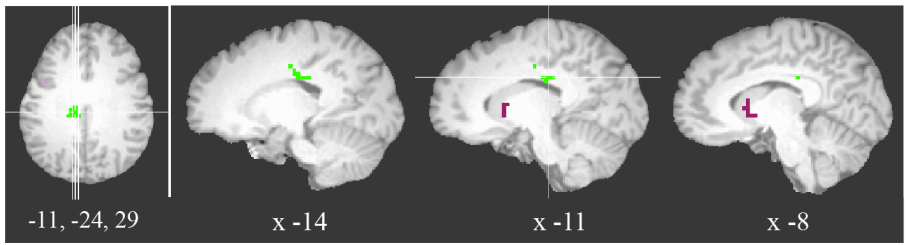
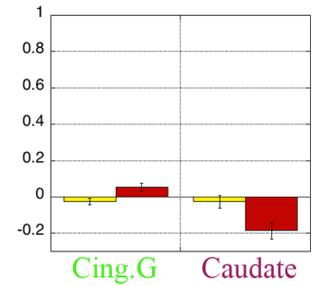

| Congruency Processing | | Timing | Semantics |
|---|---|---|---|
| (AVC - AVSI) | AVC | ✓ | ✓ |
| | AVSI | ✗ | ✗ |

ROI Means
(% signal change)

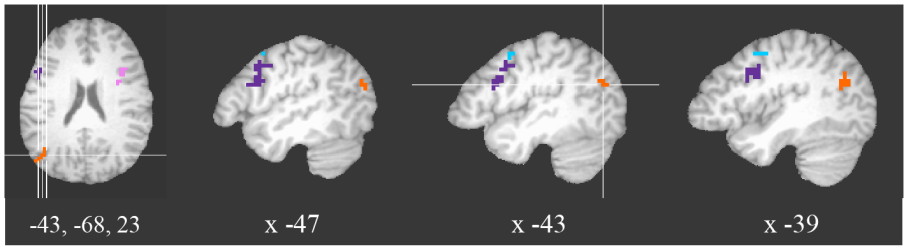
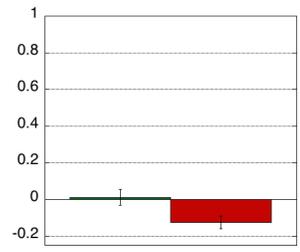

L MTG

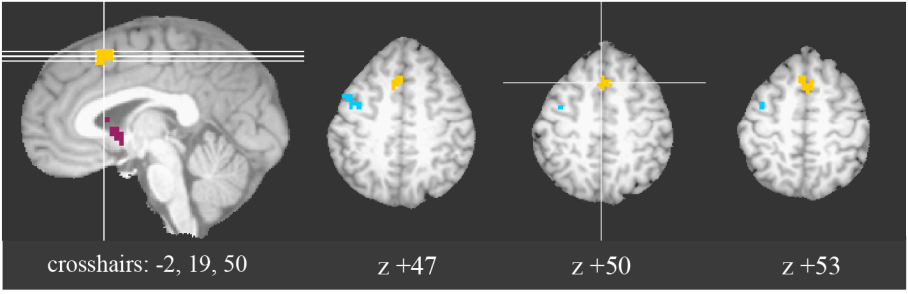
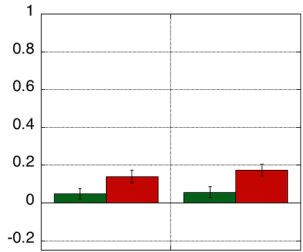

L MFG   Medial SFG

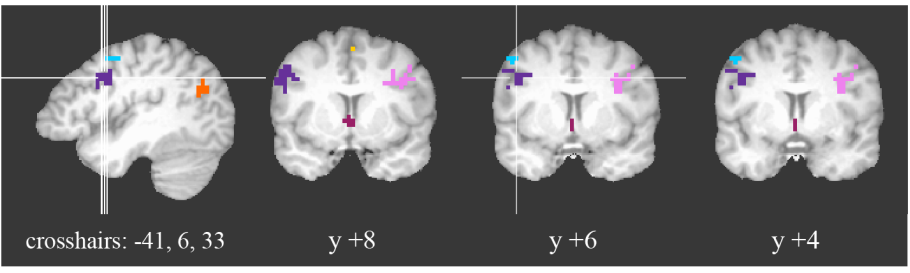
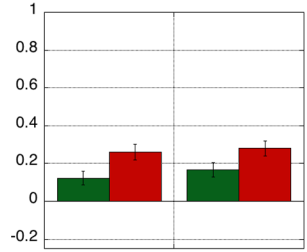

L IFG/MFG   R IFG/MFG

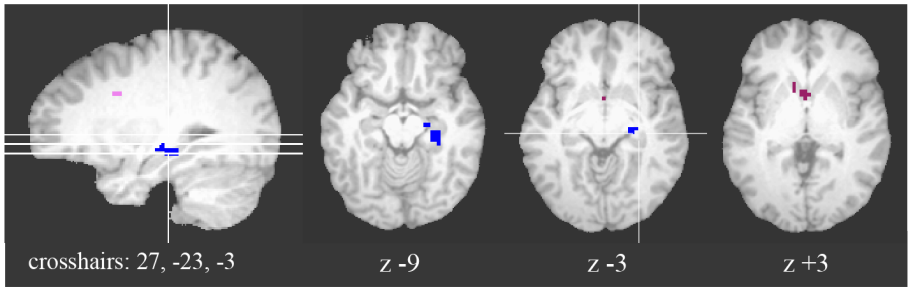
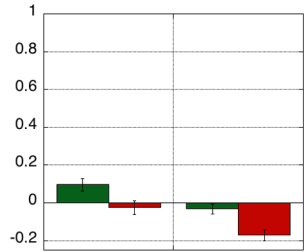

Parahipp.   Caudate

**Environmental Events**

*Below is a list of the 20 everyday events were used in this study.*

bell ringing
bouncing ball
broom sweeping
coin dropping
crumpling paper
cutting with knife
door knocking
keys jingling
knife sharpening
pen tapping
pouring water
sawing wood
shaking a waterbottle
shuffling cards
splashing water
spraying a water bottle
stapling paper
tapping silverware
tearing paper
typing on a keyboard